\documentclass[conference]{IEEEtran}
\IEEEoverridecommandlockouts
\usepackage{cite}
\usepackage{amsmath,amssymb,amsfonts}
\usepackage{algorithmic}
\usepackage{graphicx}
\usepackage{textcomp}
\usepackage{xcolor, soul}
\usepackage{enumitem}
\usepackage{subcaption}
\usepackage[normalem]{ulem}

\def\BibTeX{{\rm B\kern-.05em{\sc i\kern-.025em b}\kern-.08em
    T\kern-.1667em\lower.7ex\hbox{E}\kern-.125emX}}
\begin{document}

\title{OMD: Orthogonal Malware Detection using Audio, Image, and Static Features\\
\thanks{\text{Acknowledgement: }This work has been supported by the ONR contract \#N68335-17-C-0048. The views expressed in this paper are the opinions of the authors and do not represent official positions of the Department of the Navy.
*Lakshmanan Nataraj and Tejaswi Nanjundaswamy performed this work while they were employed at Mayachitra. }
}

\makeatletter
\newcommand{\linebreakand}{%
  \end{@IEEEauthorhalign}
  \hfill\mbox{}\par
  \mbox{}\hfill\begin{@IEEEauthorhalign}
}
\makeatother

\author{
    \IEEEauthorblockN{Lakshmanan Nataraj*}
    \IEEEauthorblockA{\textit{Mayachitra, Inc.} \\
        Santa Barbara, California \\
        lakshmanan\_nataraj@ece.ucsb.edu}
    \and
    \IEEEauthorblockN{Tajuddin Manhar Mohammed}
    \IEEEauthorblockA{\textit{Mayachitra, Inc.} \\
        Santa Barbara, California \\
        mohammed@mayachitra.com}
    \and
    \IEEEauthorblockN{Tejaswi Nanjundaswamy*}
    \IEEEauthorblockA{\textit{Mayachitra Inc.} \\
        Santa Barbara, California \\
       tejaswi@ece.ucsb.edu
       } 
    \linebreakand % <------------- \and with a line-break
    %\linebreakand
    \IEEEauthorblockN{Satish Chikkagoudar}
    \IEEEauthorblockA{\textit{U.S. Naval Research Laboratory} \\
        Washington, D.C. \\
        satish.chikkagoudar@nrl.navy.mil}
    \and
    \IEEEauthorblockN{Shivkumar Chandrasekaran}
    \IEEEauthorblockA{\textit{ Mayachitra, Inc. \& UC Santa Barbara} \\
        % \textit{ECE Department, UC Santa Barbara}\\
        Santa Barbara, California \\
        shiv@ucsb.edu}
    \and
    \IEEEauthorblockN{B.S. Manjunath}
    \IEEEauthorblockA{\textit{Mayachitra, Inc. \& UC Santa Barbara} \\
        % \textit{ECE Department, UC Santa Barbara}\\
        Santa Barbara, California \\
        manj@ucsb.edu}
}

\maketitle

%%%%%%%%%%%
% todos
% change Feature to Feature Type in Page 1 - DONE
% Fix Figure 2 - DONE
% Add a sentence about JFS metric earlier as suggested by Reviewer - ND
% add refs by rev - NA
% more specs?
% error overlap numbers in tab 1? - DONE
% fix MFCC dim as suggested - DONE
% fix fig 3 and fig 4 - DONE
% re-check jfs matrix - DONE

%%%%%%%%%%%

\begin{abstract}

% We propose a novel approach to detect and classify malware using audio descriptors. 
% We first show how audio descriptors are effective in classifying malware.
% We then show that these audio features are orthogonal to other features that have been applied to malware classification such as image descriptors, n-grams and fuzzy hashing based methods. 
%Experimental results on two malware datasets show that our approach is effective in classifying malware.
% \bsm{This sentence is quite weak; what is effective means? can you strengthen this somewhat, by making reference to SOTA or quantifying it in someway? Abstract overall can be written better.}
% We also demonstrate the orthogonality of different features and show which features contribute more towards classifying malware. 

With the growing number of malware and cyber attacks, there is a need for ``orthogonal'' cyber defense approaches, which are complementary to existing methods by detecting unique malware samples that are not predicted by other methods. 
In this paper, we propose a novel and orthogonal malware detection (OMD) approach to identify malware using a combination of audio descriptors, image similarity descriptors and other static/statistical features.
First, we show how audio descriptors are effective in classifying malware families when the malware binaries are represented as audio signals.
Then, we show that the predictions made on the audio descriptors are orthogonal to the predictions made on image similarity descriptors and other static features.
Further, we develop a framework for error analysis and a metric to quantify how orthogonal a new feature set (or type) is with respect to other feature sets.
This allows us to add new features and detection methods to our overall framework. 
Experimental results on malware datasets show that our approach provides a robust framework for orthogonal malware detection. %to detect malware and demonstrates the orthogonality of different malware detection features and show which features contribute more towards classifying malware. 

% While this binary/behavioral signatures battle front is being fought, it may be beneficial for defender to open several more cyber battle fronts to make it more expensive for the adversary to develop successful/undetected exploits. A new cyber battle front implies that it employs new detection vectors which is/are orthogonal (independent) to the current techniques of binary/behavioral signature based detections, such as [1,2,3,4].
%We are hoping that these novel orthogonal detection techniques can raise the difficulty factor and cost for successfully developing and deploying an exploit or malware by requiring attackers to contend with many distinct and orthogonal detection vectors, multiplying their cost.
%Orthogonal detections can help reduce the sheer number of malwares and exploits targeted toward our military networked computing systems.

\end{abstract}


\vspace{-15pt}

%%%%%%%%%%%%%%%%%%%%%%%%%%%%%%%%%%%%%%%%%%%%%%%%%%%%%%%%%%%%%%%%%%%%%%%%%%

%% todos
% reduce decimals and fit to single column
% bar graphs
% spectrograms
% reduce refs to 20+

% mention altohugh, we show our results on static features, our framework is easily extendable to dynamic and other features as long as the features are orthogonal

%%%%%%%%%%%%%%%%%%%%%%%%%%%%%%%%%%%%%%%%%%%%%%%%%%%%%%%%%%%%%%%%%%%%%%%%%%

\section{Introduction}
% A malware binary is read as a one dimensional signal of 8-bit unsigned integers, where every entryof is a byte value of the malware. 
% The range of this signal is [0,255] (0: black, 255: white).  Figure 1shows an example of representing a malware binary as a digital signal.  In this project we treat thisdigital signal as an 8-bit audio signal.
% \sout{{TODO1: "feature" TO "feature type"}}

% \sout{{TODO2: Observation 4 - Sensitivity to class imbalance}}

Despite increasing efforts in computer security, malware continues to be a large problem for everyone from private individuals to national governments.  
With the significant number of new malware being generated, it is difficult for traditional methods to keep up.
Today's commercial Antivirus defense mechanisms are based on scanning computers for suspicious activity. 
If such an activity is found, the suspected files are quarantined and the vulnerable system is patched with an update. %which can a significant downtime.
In turn, the Antivirus software are also updated with new signatures to identify such activities in future.
These scanning methods are based on a variety of techniques such as static analysis, dynamic analysis, and other heuristics based techniques, which are often slow to react to new attacks and threats. While this signatures battle front is being fought, it may be beneficial for the defender to open several more cyber battle fronts to make it more expensive for an adversary to develop successful/undetected exploits.
A new cyber battle front implies that it employs new detection vectors which are ``orthogonal'' (independent) to the already existing malware detection methods, i.e., a new detection method correctly identifies malware that are not detected by existing methods. 
%current signature based detection methods.
In this way, orthogonal detections can help reduce the large number of malware and exploits targeted towards military networked computing systems.

\begin{figure}[t]
\begin{center}
\includegraphics[width=0.95\columnwidth]{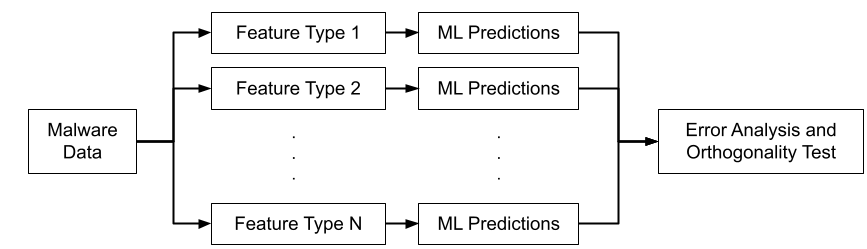}
\end{center}
\vspace{-5pt}
\caption{Overall framework of OMD - A malware detection system where multiple input feature types (static, statistical, image, audio) are passed through Machine Learning (ML) classifiers. The predictions from these classifiers are then checked to see if the features are orthogonal.}
\vspace{-15pt}
\label{fig:ortho-framework}
\end{figure}

\begin{figure*}[t]
	\centering
	\includegraphics[trim={0 4.15cm 0 0},clip,width=0.4\textwidth]{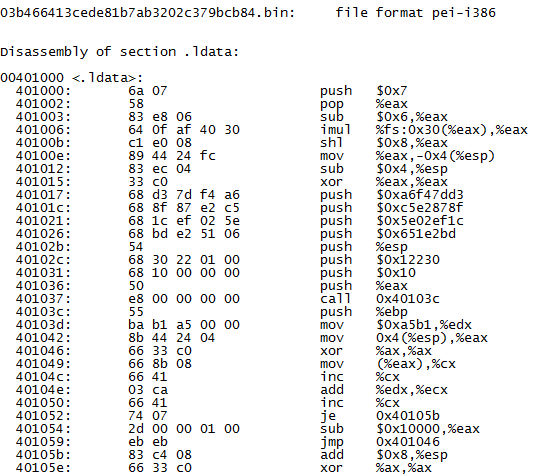}
	\includegraphics[trim={0 4.15cm 0 0},clip,width=0.4\textwidth]{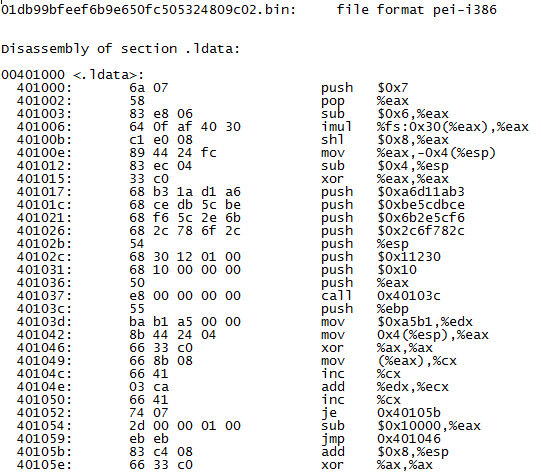}
	\caption{Disassembly of two variants from \emph{Alueron} family where most of the instructions between the two variants are same.}
	\label{fig:alueron-objdump}
	\vspace{-15pt}
\end{figure*}

%%%%%%%%%%%%%%%%%%%%%%%%%%%%%%%%%%%%%%%%%%%%%%%%%%%

In prior work, we have demonstrated effective automated malware classification by recasting software binaries as signals/images and exploiting computer vision techniques ~\cite{nataraj2011malware,nataraj2016spam}.
In this paper, we extend this approach by representing malware binaries as audio signals and computing audio descriptors to classify malware.
Then, we show how the predictions made on audio descriptors are orthogonal to predictions made on other methods like image similarity features and static analysis features. 
%By orthogonal approaches, we mean cyber defense approaches which are complementary to existing methods and identify unique malware samples that are not detected by other methods.
%\bsm{refer to SPAM paper. We undervalue our own references frequently, and that opens up opportunities for others to cite non-original work.}
% When malware binaries are visualized as digital images, the images belonging to the same malware family appear similar in layout and texture. We exploit this visual similarity and developed techniques for analyzing malware.  
% Here, we developed a proof of concept prototype  to  detect  malware  based  on  visual and signal  similarity.
% We  identified  the  metrics  that determine  the  prototype’s  efficacy  in  terms  of  accuracy  and  speed.   We  use  global  signal and image similarity descriptors that compactly summarize an image.  Typically images are resized to a standard size before computing these descriptors. Our results demonstrate that the amount of resizing plays an effective role in determining the accuracy of malware detection. Next we investigated the orthogonality of image similarity descriptors with the current generation of malware detection  
% \bsm{this is confusing, why are you referring to image descriptors while the paper is about audio descriptrors?}.
Since malware authors are usually familiar with standard detection methods and produce new malware that evades these techniques, the hope is that these orthogonal methods raise the difficulty factor and cost for successfully developing and deploying new malware. 
% To this end, our experiments demonstrate that signal and image similarity descriptors can identify malware that are not detected by state-of-the-art detection techniques.
We demonstrate our orthogonal malware detection (OMD) framework on standard static features based on code disassembly, n-grams, as well as signal processing based features such as image and audio descriptors. 
We focus on static and signal processing methods as a proof of concept, but our framework can be easily extended to other methods like dynamic analysis as long as these methods are orthogonal.
The overall framework of our proposed OMD approach is shown in Figure~\ref{fig:ortho-framework}.
The main contributions of this paper are:
\begin{itemize}[leftmargin=*]
\item We first show how the similarity of malware variants when represented as audio signals can be leveraged for identifying malware families. We then perform a thorough investigation of audio descriptors for malware classification and detection. 
\item We also show how the predictions based on audio descriptors are orthogonal to predictions based on other static and statistical features. 
\item Finally, we layout a robust and practical framework (OMD) that can be used by defense/military for testing a malware detection system with multiple input features. By providing an orthogonality metric and error analysis of the different input feature sets, our framework allows an analyst to decide which features are best to make the cyber-defense complex. Furthermore, our framework is flexible that new feature sets can be easily added or removed if needed.
\end{itemize}
% The rest of the paper is organized as follows. 

% main contribs
% audio
% orthogonality
% robust framwork in a military/defense setting which leaves room to add new features at the discretion of an analyst. 

%%%%%%%%%%%%%%%%%%%%%%%%%%%%%%%%%%%%%%%%%%%%%%%%%%%%

% \vspace{-8pt}
\section{Related Work}
\label{s:rw}

%\vspace*{-0.15in}
% static, dynamic, statistical
% recent deep learning based

Typical features extracted from malware can be broadly grouped into either \emph{static} or \emph{dynamic} features.
As the name suggests, static features are extracted from the malware without executing it. 
Dynamic features, on the other hand, are extracted by executing the sample, usually in a virtual environment, and then studying their behavioral characteristics such as system calls trace or network behavior. 
%Among dynamic analysis, the most common method is to execute the malware in a controlled environment and then study its execution behavior. 
Recently, there has been lot of research on malware detection using deep learning~\cite{vinayakumar2019robust,gibert2020rise,mohammed2021malware}.
%, recurrent neural networks (RNNs)~\cite{rhode2017early,shukla2019rnn,jha2020recurrent}, convolutional neural networks (CNNs)~\cite{kim2017image,vasan2020image,} and hybrid models~\cite{mishra2019vmanalyzer,sajid2020dodgetron,wang2020botmark}.
%The works that use CNNs, typically converts malware binaries to digital images and pass them into a CNN in order to detect malware.
% However, in this paper, we visualize malware in the frequency domain and then use CNNs for malware detection. 
%\bsmres{what is visualization to do with analytics? focus on analysis, visualization is a by-product of explaining what you are doing. - REMOVED}
%The proposed malware detection approach also has all the positives of static analysis methods while addressing some of the limitations of the aforementioned works like high time complexity, low scalability and high total feature selection count.
%We will review some of the approaches here.
% Here, we will mainly review works that focus malware classification and detection based on similarity between malware.

% \vspace{-10pt}

First, we will briefly review the static analysis and signal processing based methods. 
Static analysis can further be classified into static code based analysis and non code based analysis.  
Static code based analysis techniques study the functioning of an executable by disassembling the executable and then extracting features.
One common static code based analysis is control flow graph analysis~\cite{nguyen2018auto,ma2019combination}.
%~\cite{ding2014control,alam2015annotated,nikolopoulos2017graph,nguyen2018auto,ma2019combination}. 
%\bsm{As I pointed out in the other paper, you have 3x time the references for a conference paper. This dilutes the overall message as well as the cited references. Why dont you select one representative/seminal or overview paper instead of citing 4-5 papers for each? Further, make sure that when it comes to novel ways of looking at the problem, cite your initial works first rather than citing some follow-up works of others.}
% carrera2004digital,bruschi2006detecting,kruegel2006polymorphic,anju2010malware,
After disassembly, the control flow of the malware is obtained from the sequence of instructions and graphs are constructed to uniquely characterize the malware.
% A graph similarity measure is used to measure the similarity between two malware.
% However, these methods do not work well on packed/obfuscated malware since the control flow of a packed malware reveals only the unpacking routine and not the actual flow.
% abou2004n,kolter2006learning,shafiq2008embedded,santos2009n,
% kornblum2006identifying,yen2008traffic,griffin2009automatic
% schultz2001data,ye2008intelligent,shafiq2009pe, wicherski2009pehash,liao2012pe,
% jacob2012static,kancherla2013image,
% kolter2006learning,
% shiel2019improving,
% biondi2019effective,
% 
Static non code based techniques are based on a variety of techniques: n-grams~\cite{zhang2019classification},  hash based techniques~\cite{namanya2020similarity}, file structure~\cite{rezaei2020efficient}  or signal similarity based techniques~\cite{nataraj2011malware,zhang2016malware,nataraj2016spam}. %mohammadi2017proposing
Recently, there have also been approaches to detect malware using audio descriptors~\cite{farrokhmanesh2016novel,sharma2018image,azab2020msic,mercaldo2021audio}, though these works do not focus on the orthogonality part. 
Below, we will review some of the above-mentioned features which we will later use in our experiments. 
%\subsection{Feature Extraction}

% \subsection{GIST Descriptor}
% \begin{figure}[t]
% 	\centering
% 	\includegraphics[width=0.4\textwidth]{figures/alueron-var-1.png}
% 	\includegraphics[width=0.4\textwidth]{figures/alueron-var-2.png}
% 	\caption{Visualizations of two variants from \emph{Alueron} family}
% 	\label{fig:alueron-vis}
% \end{figure}
\noindent \textbf{GIST Descriptor:} In prior work, we have demonstrated that visual similarity among malware variants of same families can be exploited, thus giving rise to image similarity descriptors to detect malware~\cite{nataraj2016spam,nataraj2011malware}. 
% In particular, we showed that GIST descriptors obtain high accuracy in detecting malware and are orthogonal to state of the art security based methods.
Briefly, to extract this descriptor the image is filtered with a Gabor filter bank of different scales and orientations (typically 8 orientations and 3 scales).
These outputs are divided into a number (e.g. 20) of subbands, with the average value computed for every subband to obtain a 320-dimensional (320-D) feature vector.
%Over large spatial regions, the magnitude of feature values are averaged for each subband and then downsampled to a fixed rectangle size, e.g. $4 \times 4$.
%See Figure~\ref{fig:gistExtract} for a visual walkthrough. 
% From the example parameters given here ($4 \times 4$ downsampled size, and 20 subbands) this would result in a $4 \times 4 \times 20 = 320$ dimensional feature vector. 
%We extract the GIST descriptors on both the malware dataset and benign files.

% \begin{figure}[t]
% \begin{center}
% \includegraphics[width=0.9\columnwidth]{figures/full_feature.png}
% \end{center}
% \vspace{-12pt}
% \caption{An example of GIST extraction, figure from \cite{nataraj12SARVAM}}
% \label{fig:gistExtract}
% \end{figure}

\begin{figure*}[t]
\begin{center}
\includegraphics[width=0.75\textwidth]{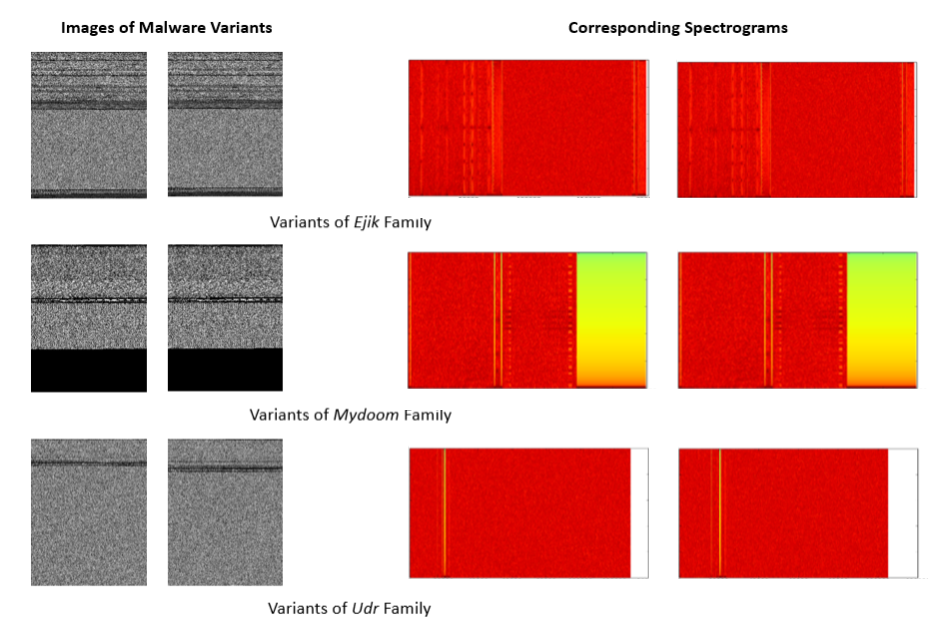}
\end{center}
\vspace{-12pt}
\caption{Visualizations of malware variants as grayscale images (left) and their corresponding audio spectrograms (right) belonging to {\it Ejik}, {\it Mydoom} and {\it Udr} malware families. We can see that the spectrograms exhibit similarity for malware variants belonging to the same family, and are different from those of other families (images and spectrograms are scaled approximately to the same size for visualization).}
\vspace{-15pt}
\label{spec-vars}
\end{figure*}

% \begin{figure*}[t]
% \centering 
% \begin{subfigure}[t]{0.4\textwidth}
% {\includegraphics[width=\textwidth,height=0.5\columnwidth]{figures/mydoom-a841250d41984f977baf6712e2b1ba3e-1.png}}  %\caption{U}}
% \vspace{-15pt}
% \caption{Spectrogram of {\it Mydoom} Variant 1}
% \end{subfigure} \ \
% \begin{subfigure}[t]{0.4\textwidth}
% {\includegraphics[width=\textwidth,height=0.5\columnwidth]{figures/mydoom-ab642e4a4af67c5e5fdf41d793d9b20b-1.png}}  %\caption{U}}
% \vspace{-15pt}
% \caption{Spectrogram of {\it Mydoom} Variant 2}
% \end{subfigure} \\
% %\\ \\
% %\caption{Variant 1}
% \begin{subfigure}[t]{0.4\textwidth}
% {\includegraphics[width=\textwidth,height=0.5\columnwidth]{figures/ejik-a8ac135690740502f6c95dd72f0491c0-1.png}}  %\caption{U}}
% \vspace{-15pt}
% \caption{Spectrogram of {\it Ejik} Variant 1}
% \end{subfigure} \ \
% \begin{subfigure}[t]{0.4\textwidth}
% {\includegraphics[width=\textwidth,height=0.5\columnwidth]{figures/ejik-a961494f4c8841aaab600917f705f70a-1.png}} 
% \vspace{-15pt}
% \caption{Spectrogram of {\it Ejik} Variant 2}
% \end{subfigure}
% \vspace{-5pt}
% \caption{Audio spectrograms of malware variants belonging to {\it Mydoom} and {\it Ejik} family. We can see that the spectrograms exhibit similarity for malware variants belonging to the same family.}
% \label{spec-vars}
% \vspace{-10pt}
% \end{figure*}

%\subsection{N-grams of Bytes}
% refer to old Data Mining Paper, koetler and maloof, aboud
\noindent \textbf{N-grams of Bytes:} n-grams of byte codes have been effective in classifying and detecting malware~\cite{kolter2006learning}.
%\bsm{pick one that is previously cited, no need for a laundry list here.}
As an example, for the byte stream  ‘0a 1b c4 8a’, the
corresponding 2-grams (bigrams) will be 0a1b, 1bc4, c48a.
We use the 2-grams frequency count to obtain a $2^{16}$ (65,536-D) feature vector.

\noindent \textbf{N-grams of Disassembled Instructions:} Apart from n-grams of bytes, n-grams of disassembled instructions are also good features to characterize malware~\cite{lakhotia2013vilo}.
%\bsm{pick one that was previously cited. No need to add an additional citation ref 55.}
We use the simple \emph{objdump} tool for disassembly.
% \st{In future, we plan to use more complex tools such as Hex-Rays IDA cite idapro to disassemble sophisticated malware. } 
% \bsm{this is not a conclusion/discussion section, only refer to what you have done for this paper. }
Figure~\ref{fig:alueron-objdump} shows the disassembly for two variants of \emph{Alueron} family.
We see that most of the instructions are similar.
We will use the instruction count of 46 most common instructions as our feature vector. 

%\subsection{pehash}
\noindent \textbf{\emph{pehash}:} This method utilizes Portable Executable (PE) file structure characteristics and computes a hash to cluster polymorphic malware variants~\cite{wicherski2009pehash}.
This is based on the assumption that polymorphic malware share the same PE file structure properties.
Attributes such as image characteristics, heap commit size, stack commit size, virtual address, raw size and section characteristics are used to compute a 40-character long hash.
For example, hash values of two variants from \emph{Adialer} malware family will be: \\ 
\textbf{Variant 1: }\emph{dbf2a2ef1fed22c6e20ffa1b9a5ab69a75c365eb}\\
\textbf{Variant 2: }\emph{dbf2a2ef1fed22c6e20ffa1b9a5ab69a75c365eb}

Since the hash value is the same, the number of collisions between different hashes in a database is used to compute similarity.
To facilitate faster processing and distance computation, we convert the characters to their ASCII value and obtain a 40-D \emph{pehash} feature vector for every sample.

\section{Methodology and Experiments}
% \vspace{-6pt}

\subsection{Similarity of malware when represented as audio signals}
\label{sec:audio-sim}

We have previously shown that malware variants exhibit visual similarity when represented as digital images~\cite{nataraj2011malware,nataraj11comparative}.
Here, a malware binary is read as a one dimensional signal of 8-bit unsigned integers, where every entry is a byte value of the malware. 
The range of this signal is $[0,255]$. % (0: black, 255: white).
%Here we explore the same with audio signals. 
In Fig~\ref{spec-vars}, we notice this similarity even in the audio signals when visualized as spectrograms (time vs frequency) for malware variants belonging to {\it Ejik}, {\it Mydoom} and {\it Udr}  families. 
These spectrograms are similar for variants within a family while different from variants of other families. %We had observed a similar phenomenon on malware variants that come from many other families too.
For comparison, the grayscale images of these samples are also shown in Fig~\ref{spec-vars}.
This motivated us to explore audio similarity descriptors for malware detection.
%%%%%%%%%SATISH
We investigated the following audio descriptors that are commonly used in audio and music analysis using the  {\it librosa} package~\cite{mcfee2015librosa}:
\begin{itemize}[leftmargin=*] 
%\addtolength{\itemindent}{2cm}
    \item Chroma: This feature projects the audio spectrum into 12 bins corresponding to distinct semitones (or chroma) within the musical octave. 
    This feature exploits the fact that humans perceive notes that are one octave apart similarly. 
    Hence understanding the characteristic of a musical piece within an octave helps identify musical similarity. The \emph{librosa} package provides 3 variations for calculating Chroma features:
    \begin{itemize}
        \item Chroma-STFT: The first option uses short term fourier transform (STFT) to get the audio spectrum. 
        \item Chroma-CQT: The second option uses constant Q-transform (CQT) \cite{schorkhuber2010constant} to get the audio spectrum. 
        \item Chroma-CENS: The third option employs an additional step of energy normalization to get chroma energy normalized statistics (CENS) \cite{muller2005audio}.
    \end{itemize}
    \item Mel Frequency Cepstral Coefficients (MFCC): This feature captures the shape of the spectral envelope in 20 coefficients to form a 20-dimensional (20-D) feature vector. The MFCC are obtained by first calculating the power spectrum from STFT, then mapping these to mel scale using triangular windows (to mimic the human auditory system), then taking log of these mel scale power coefficients, and finally taking the discrete cosine transform of these log mel power coefficients. The MFCC are popular features for speech recognition as the spectral envelope shape can distinguish between different phenoms of speech~\cite{rabiner1993fun}.
    \item Melspectrogram: This feature is similar to MFCC, wherein the magnitude spectrum is first calculated using STFT and then mapped to mel scale using 128 triangular windows. 
    These coefficients capture more detailed frequency information compared to only spectral envelope shape of MFCC.
\end{itemize}

\subsubsection{Preliminary experiments using MFCC features}
\label{sec:ortho}

As a preliminary experiment, we used the Malimg dataset which contains variants from 25 malware families~\cite{nataraj2011malware}.
The dataset has a mixture of both packed and unpacked malware and the number of samples per family varies from 80 to 2,949.
% There are a couple of motivations to use this dataset as a starting point in our testing. 
% The size of the dataset, 9339 malware samples from 25 families, is large enough to not be trivial, but small enough for quick experimentation, as a complete feature extraction can be done in a few hours, and search time is very quick.
% For the benign files, we obtained the hashes of Windows system executables~\cite{sarvamblog-packed-benign} belonging to various Windows OS (XP, Vista, NT, 7) and crawled them from SARVAM~\cite{SARVAMsite}.
%\subsubsection{Experimental Setup and Results}
We first convert them to audio signals and compute MFCC features. Using 10-fold cross validation, we divide the data into 10 disjoint folds and chose 9 folds for training, 1 fold for testing and then vary the folds for 10 iterations.
The classification accuracy is then obtained for each iteration and then averaged over all iterations.
The Nearest Neighbor classifier is used for classification.
For this experiment, we obtained a high classification accuracy of 97\%, thus showing that audio descriptors are able to correctly classify most families as shown in Figure~\ref{fig:mfcc-conf}.
%\subsubsection{Results}
%The results of our experiments are shown in Figure~\ref{fig:mfcc-conf}. 
%From the confusion matrix, we see that the audio descriptors are able to correctly classify most families.
% The confusion among few families are those which have very similar characteristics.
% Grouping them into a single family could eliminate the confusion.

%%%%%%%%%%%%%%%%%%%%%%%%%%%%%%%%%
\begin{figure}[t]
 \begin{center}
 \includegraphics[width=0.68\columnwidth,height=0.55\columnwidth]{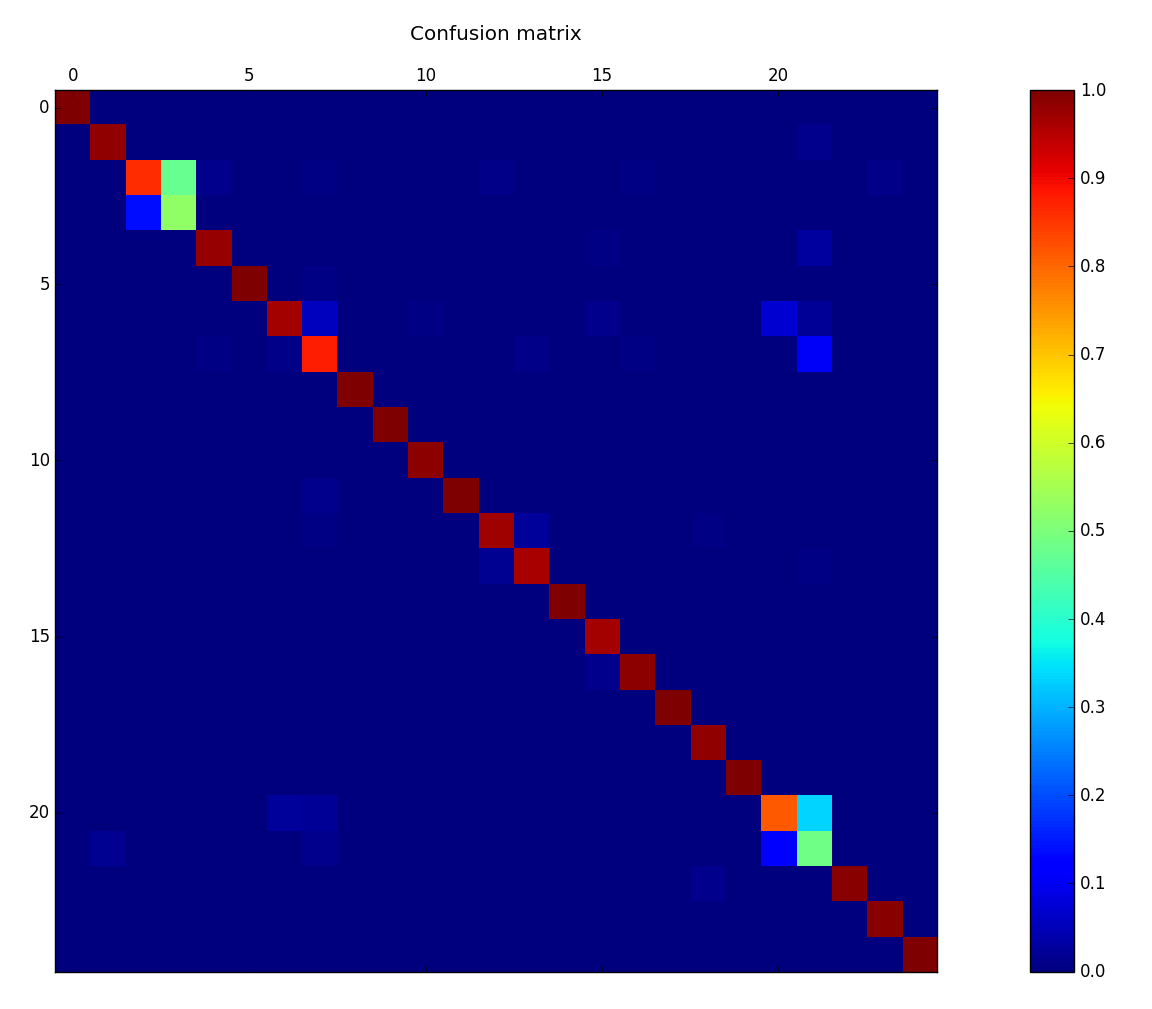}
 \end{center}
 \vspace{-12pt}
 \caption{Confusion matrix for malware classification of different families using MFCC features.}
  \label{fig:mfcc-conf}
  \vspace{-18pt}
 \end{figure}

%%%%%%%%%%%%%%%%%%%%%%%%%%%%%%%

% \subsubsection{Preliminary Experiments using MFCC Features}
% \label{sec:ortho}
%\subsubsection{Orthogonaity Test using MFCC Features:}

We also setup another experiment where the objective is to verify if MFCC features (20-D) can identify malware samples from benign samples.
Here, we append the Malimg dataset with benign system files from different versions of Windows Operating systems. 
In this way, we generated a \emph{small dataset} with 9,130 samples from MalImg dataset~\cite{nataraj2011malware} + 7,228 benign samples from Windows system files.
To evaluate orthogonality, we perform a comparative assessment of MFCC features with various feature sets previously used for malware detection:
\begin{itemize}[leftmargin=*] 
%\addtolength{\itemindent}{2cm}
   \item GIST descriptors based on image similarity (320-D)
    \item n-grams of byte codes (65,536-D)
    \item Assembly Language Instruction count (46-D)
    \item {\it pehash} (128-D)
    %\item {\it ssdeep} (100-D)
\end{itemize}
%We briefly describe these features below.

%%%%%%%%%%%%%%%
%\subsubsection{Experimental Results}
Using Nearest Neighbor classifier, we compare the classification accuracy of all the feature sets using 10-fold cross validation.
The results are shown in Table~\ref{tab:acc-comp}. %Figure~\ref{fig:malvsben-res}.
We see that the accuracy of MFCC features is comparable with n-grams, instruction count and GIST but outperforms \emph{pehash}. % , demonstrating some similarities and differences with other approaches. 
%\bsm{the following sentence can be reworded better. Here is an attempt: We note that the 2-dim MFCC features are very compact, compared to n-grams that is 3000 times larger at 65K dimensions, or the 320-dim GIST features. }
We note that the 20-D MFCC features are very compact, compared to n-grams that is 3,000 times larger at 65,536-D, or the 320-D GIST features. %, thus making it highly compact.
% We note that the dimensionality of MFCC features (20-D) is {\bf 3276} times lesser than n-grams (65536-D), {\bf 16} times less than GIST (320-D) and {\bf 2} times less than assembly language instruction (46-D) features, thus making it highly compact. 
%\bsm{with regard to instruction count, it is comparable. Your following statement makes that point, so you may want to separate the compactness part for the instruction count to rephrase it together with the challenges in disassembly.}
We do not consider the byte n-grams feature in our future experiments due to it's large computational footprint.
For 46-D instruction count feature, though the 20-D MFCC feature is comparable in size, the binary needs to be disassembled.
This is challenging since malware are known to be designed with anti-disassembly mechanisms~\cite{branco2012scientific} to evade reverse engineering.
On our dataset 209 malware could not be disassembled.
But this problem will not arise for MFCC features since it directly works on the bytes.
The concise representation of MFCC features, while maintaining similar/superior performance, points to a difference in underlying information used, and thus shows orthogonality.
% \begin{figure}[!htbp]
% 	\centering
% 	\includegraphics[width=0.28\textwidth,height=0.4\columnwidth]{figures/mal-vs-ben-result-aug-2016.png}
% 	\caption{Comparative evaluation of Features for Malware Detection. The accuracy of MFCC is almost the same as that of n-grams, instruction count and GIST while outperforming hash based \emph{pehash}.}
% 	\label{fig:malvsben-res}
% 	\vspace{-10pt}
% \end{figure}

\begin{table}[h]
\centering
\caption{Comparison of Accuracy of MFCC features.}
\vspace{-5pt}
\begin{tabular}{|c|c|c|c|c|} \hline
\bf n-grams  & \bf ins count  & \bf pehash & \bf gist & \bf MFCC\\ \hline
0.9937 & \textbf{0.9964} & 0.9461 & 0.9957 & 0.9887  \\
\hline
\end{tabular}

\label{tab:acc-comp}
\vspace{-10pt}
\end{table}

% ngrams errors: 33 
% inscnt errors: 23
% pehash errors: 1397
% gist errors: 49
% mfcc errors: 

\subsubsection{Error analysis showing signs of orthogonality}
To further evaluate orthogonality, we performed an error analysis to see if MFCC features can correctly identify samples that were misclassified by other feature sets (and vice-versa). 
% In Table~\ref{ortho-error}, we see that out of 15,920 samples, MFCC features correctly classified {23} samples against n-grams, {18} against instruction count, {43} samples against GIST and {1,388} against \emph{pehash}.
% ABOVE NUMBERS WERE SLIGHTLY OFF. REPLACING WITH CORRECT NUMBERS (DATE: 10/15/2021)
The number of incorrectly classified samples (errors) were {33} for n-grams, {23} for instruction count, {49} for GIST and {68} for MFCC features.
Among these samples, MFCC features had an error overlap of {5} samples with n-grams, {4} with instruction count, {1,397} with \emph{pehash} and {6} with GIST features.
This shows that MFCC features correctly identified {28} samples (33-5) that were misclassified by n-grams, while n-grams correctly identified 63 samples (68-5) that were misclassified by MFCC features, thus establishing that these two feature sets can make predicitons that are orthogonal to each other.
Thus, in Table~\ref{ortho-error}, we see that MFCC features correctly classified {28} samples against n-grams, {19} against instruction count, {43} samples against GIST and {1,374} against \emph{pehash}.
This strengthens the argument of the orthogonality of MFCC features.
%out of 15,920 samples

% old result
% \begin{table}[h]
% \centering
% \caption{Samples correctly classified by Gist but misclassified by other features and vice versa}
% \label{ortho-error}
% \begin{tabular}{|c|c|c|c|} \hline
%  & n-grams  & ins count  & \emph{pehash} \\ \hline
% Gist correct, Others wrong & 25 & 21 & 1368   \\ \hline
% Gist wrong, Others correct & 40 & 46 & 36  \\ \hline
% \end{tabular}
% \end{table}

%\bf MFCC Error Overlap & 5 & 4 & 1,397 & 6 \\ 
%\hline

\begin{table}[h]
\centering
\caption{Samples correctly classified by MFCC features but misclassified by other feature sets and vice versa.}
\vspace{-5pt}
\begin{tabular}{|c|c|c|c|c|} \hline
 & \bf n-grams  & \bf ins count  & \bf pehash & \bf gist\\ \hline
%\bf MFCC $\checkmark$, Others $\times$ & 23 & 18 & 1,388 & 43  \\ \hline
% % ABOVE NUMBERS WERE SLIGHTLY OFF. REPLACING WITH CORRECT NUMBERS (DATE: 10/15/2021)
%\bf MFCC Error Overlap & 5 & 4 & 1,397 & 6 \\ 
%\hline
\bf MFCC $\checkmark$, Others $\times$ & 28 & 19 & 1,374 & 43  \\ \hline
\bf MFCC $\times$, Others $\checkmark$ & 63 & 64 & 45 & 62  \\ \hline
\end{tabular}

\label{ortho-error}
\vspace{-12pt}
\end{table}

\subsection{Orthogonality metric}

To further understand these results and to measure their orthogonality with respect to other feature sets, we use a metric called the ``Joint Feature Score ($JFS$)'' that we had previously used in~\cite{mohammed2021malware}. % and covered in Report 4 (submitted on November 1st 2017).
%We will review the metric here.
To calculate $JFS$, we first calculate the error-analysis matrix, $E$, that compares classification errors for different feature sets. Specifically, $i$th row in this matrix corresponds to one feature set-$i$, and the $j$th column in this row corresponds to the number of samples that were misclassified by the feature set-$i$ but were correctly classified by feature set-$j$. 
We then normalize row $i$ by the number of samples misclassified by feature set-$i$ to generate normalized error-analysis matrix, $E_N$. 
If two feature sets $i,j$ ($i \neq j$) are orthogonal, i.e., errors using feature set-$i$ are completely different from those for feature set-$j$, then the normalized error matrix will have entries `1' in for $(i,j)$ and $(j,i)$. 
Using this criteria we define the ($JFS$) between feature sets $i$ and $j$ as
\begin{equation}
JFS(i,j) = (2 - \left( ( 1-E_N(i,j) )^2 + ( 1-E_N(j,i) )^2 \right)^{1/2} )/2
\end{equation}
Clearly, if the feature sets are orthogonal, $JFS$ will be 1, and if their errors overlap completely, then it will be 0.

\begin{table}[t]
\centering
\caption{Classification accuracies when using audio descriptors on the small dataset.}
\vspace{-5pt}
\begin{tabular}{|c|c|c|c|c|}
\hline
\textbf{}                        & \textbf{k-NN} & \textbf{RF} & \textbf{MLP} & \textbf{AdaBoost}\\ \hline
\textbf{MFCCe (320d)}     & \textbf{0.9932} &	0.9931 & \textbf{0.9963} & 0.9873 \\ \hline
\textbf{Melspectrogram (384d)}   & 0.9786 &	\textbf{0.9961} & 0.9953 & \textbf{0.9929} \\ \hline
\textbf{Chroma-STFT (312d)}      & 0.9550 &	0.9761 & 0.9663 & 0.9506  \\ \hline
\textbf{Chroma-CQT (312d)}        & 0.9196 & 0.9370 & 0.9560 & 0.8798  \\ \hline
\textbf{Chroma-CENS (312d)}        & 0.8852 & 0.9360 & 0.9416 &	0.8640  \\ \hline
\end{tabular}

\label{tab:exp-dim}
% \vspace{-8pt}
\end{table}

\begin{table}[t]
\centering
\caption{Number of classification errors for different feature sets on the small dataset.}
\vspace{-5pt}
\begin{tabular}{|c|c|c|}
\hline
\textbf{Features}                        & \textbf{Accuracy} & \textbf{\# incorrectly classified samples} \\ 
   &   & \textbf{(out of 1,636)} \\ 
\hline
\textbf{gist (320d)}     & 0.9951 &	  8      \\ \hline
\textbf{instrcount (46d)}     & \textbf{0.9969} &	  5       \\ \hline
\textbf{pehash (40d)}     & 0.9113 &	  145     \\ \hline
%\textbf{ssdeep (40d)}     & 0.6381 &	  592       \\ \hline
%\textbf{sdhash (20d)}     & 0.7249 &	  450       \\ \hline
%\textbf{MFCC (385d)}     & 0.9890 &	  18       \\ \hline
% \textbf{Melspectrogram (128d)}   & 0.9835 &  27	        \\ \hline
% \textbf{Chroma-STFT (12d)}        & 0.9303 & 114	       \\ \hline
% \textbf{Chroma-CQT (12d)}        & 0.8704 &  212	       \\ \hline
% \textbf{Chroma-CENS (12d)}        & 0.8478 & 249	       \\ \hline
\textbf{MFCCe (320d)}     & 0.9932 & 11	        \\ \hline
% \textbf{Melspectrogram-expanded (384d)}   & 0.9834963325&   27	        \\ \hline
% \textbf{Chroma-STFT-expanded (312d)}        & 0.9492665036& 83	       \\ \hline
% \textbf{Chroma-CQT-expanded (312d)}        & 0.9278728606&  118	       \\ \hline
% \textbf{Chroma-CENS-expanded (312d)}        & 0.8863080684& 186	       \\ \hline
\end{tabular}

\label{tbl:error_numbers}
\vspace{-20pt}
\end{table}

\begin{table}[b]%[!htbp]
\caption{Error analysis for feature sets on the small dataset.}
\vspace{-5pt}
\centering
% \tiny
% \setlength\tabcolsep{.5pt} % default value: 6pt
\begin{tabular}{|c|c|c|c|c|}
\hline
\textbf{Error}  & \textbf{gist} & \textbf{instrcount} & \textbf{pehash} & \textbf{MFCCe} \\ 
\hline
\textbf{gist}     		        & -	  & 8   & 6	 & 6   \\ \hline
\textbf{instrcount}     	    & 5	  & -   & 2	 & 5    \\ \hline
\textbf{pehash}     		    & 143 & 142 & -	 & 144   \\ \hline
\textbf{MFCCe}     	            & 9   & 11  & 10	 & -	   \\ \hline
\end{tabular}
\label{tbl:ea_mat}
% \vspace{-5pt}
\end{table}

% \begin{table}[b]%[!htbp]
% \centering
% % \tiny
% % \setlength\tabcolsep{.5pt} % default value: 6pt
% \begin{tabular}{|c|c|c|c|c|}
% \hline
% \textbf{Error}  & \textbf{gist} & \textbf{instrcount} & \textbf{pehash} & \textbf{MFCCe} \\ 
% \hline
% \textbf{gist}     		        & 0	  & 1   & 0.75	 & 0.75   \\ \hline % div by 8
% \textbf{instrcount}     	    & 1	  & 0   & 0.4	 & 1    \\ \hline % div by 5
% \textbf{pehash}     		    & 0.986 & 0.98 & 0	 & 0.993  \\ \hline % div by 145
% \textbf{MFCCe}     	            & 0.818   & 1  & 0.909	 & 0	   \\ \hline % div by 11
% \end{tabular}
% \caption{Normalized Error analysis for features on the small dataset.}
% \label{tbl:en_mat}
% \vspace{-5pt}
% \end{table}

\begin{table}[b]%[!htbp]
\caption{$JFS$ score matrix comparing orthogonality of different feature sets on the small dataset -- normalization factors are derived from the last column of Table~\ref{tbl:error_numbers}.}
\vspace{-5pt}
\centering
%\tiny
%\setlength\tabcolsep{.5pt} % default value: 6pt
\begin{tabular}{|c|c|c|c|c|}
\hline
\textbf{$JFS$ Score}  & \textbf{gist} & \textbf{instrcount} & \textbf{pehash} &  \textbf{MFCCe} \\ \hline
\textbf{gist}     		        & - & 1 & 0.88 & 0.85 \\ \hline
\textbf{instrcount}     	    & 1 & - & 0.7  & 1     \\ \hline
\textbf{pehash}     		    & 0.88 & 0.7 & - & 0.95  \\ \hline
\textbf{MFCCe}     	            & 0.85  & 1  & 0.95 &  -  \\ \hline
\end{tabular}
\label{tbl:JFS_mat}
\vspace{-5pt}
\end{table}

%%%%%%
% Orthogonality MEtric
% Experimental Results on Small Dataset
%%%%
\subsection{Experimental results using extended audio features}
Here, we further extend the audio features by averaging the features for uniform divided parts of the binary files.
This allows us to capture local variations in features across the file, which enabled better classification accuracy than the non-extended features.
The expansion in features is chosen such that the number of dimensions is close to the GIST image feature of 320. Specifically, for MFCC we divide the binary file into 16 equal parts to get $16\times 20=320$ values, and refer to these extended features as MFCCe.
%For MFCC features, there was increase in accuracy of 0.2\% using the extended features. 
Similarly, for Melspectrogram we divide into 3 parts to get $128\times 3=384$ values; and for chroma we divide into 26 parts to get $12\times 26=312$ values. 
We pass these features to 4 different ML classifiers of k-nearest neighbor (k-NN), random forest (RF), deep neural network based machine learning predictor (MLP), and AdaBoost.
The experimental results (with one byte per sample) on the small dataset with 10-fold cross validation are presented in Table~\ref{tab:exp-dim}. 
%The experiments were conducted with  and all the results are obtained using 10-fold cross validation.
Clearly, increased feature dimensions improves classification accuracy. 
Particularly performance using MFCCe is now close to that obtained using image GIST feature.
These are encouraging results given that we have previously observed that classification using these two feature sets are quite orthogonal to each other.
The errors corresponding to different feature sets with k-NN classifier (k=1) are listed in Table~\ref{tbl:error_numbers}.
The error analysis matrix ($E$) for these feature sets is shown in Table~\ref{tbl:ea_mat}.
The $JFS$ score for each pair of feature sets is shown in Table~\ref{tbl:JFS_mat}.
Clearly, these scores indicate that audio features are orthogonal to other feature types. 
Next, we validate our approach on a larger dataset to see if our approach is scalable.

%%%%%%%%%%%%%%%%%%%%%%%%%%%%%%%%%%%%%%%%%%%%%%%%
% aug 2019

% \section{Enhanced malware analysis using audio descriptors}
% \label{sec:audio}

% We continued our effort on analyzing malware by representing the binary file as one dimensional signal and extracting audio features. The main audio descriptors we investigated include:
% \begin{enumerate}
% \item Chroma, which represents distinct semitones/chroma with the musical octave, in three variants based on how frequency information is calculated.
% \item Mel Frequency Cepstrum Coefficients (MFCC), which captures the spectral envelope information
% \item Mel Spectrum, which captures frequency information more inline with human auditory system
% \end{enumerate}

% \section*{Continued experiments in May - Aug 2019}
%%%
\begin{table}[t]
\centering
\caption{Classification accuracies when using audio descriptors on the MaleX dataset.}
\vspace{-5pt}
% \vspace{-5pt}
\begin{tabular}{|c|c|c|c|c|}
\hline
\textbf{}                        & \textbf{k-NN} & \textbf{RF} & \textbf{MLP} & \textbf{AdaBoost}\\ \hline
\textbf{MFCCe (320d)}     & 0.9032 &	\textbf{0.9121} & 0.8909 & \textbf{0.8477} \\ \hline
\textbf{Melspectrogram (384d)}   & \textbf{0.9026} &	0.9034 & \textbf{0.9016} & 0.8407 \\ \hline
\textbf{Chroma-STFT (312d)}       & 0.8424 & 0.8858 & 0.8561 & 0.8422 \\ \hline
\textbf{Chroma-CQT (312d)}        & 0.8155 & 0.8786 & 0.8592 & 0.8401  \\ \hline
\textbf{Chroma-CENS (312d)}       & 0.8423 & 0.8776 & 0.8621 & 0.8391   \\ \hline
\end{tabular}
\label{table:exp-dim}
% \vspace{-5pt}
\end{table}

\begin{table}[t]
\centering
\caption{Number of classification errors for different feature sets on MaleX.}
\vspace{-5pt}
\begin{tabular}{|c|c|c|}
\hline
\textbf{Features}   & \textbf{Accuracy} & \textbf{\# incorrectly classified samples} \\ 
   &   & \textbf{(out of 104,440)} \\ 
\hline
\textbf{gist (320d)}                        & 0.8888 &	  11,611      \\ \hline
\textbf{instrcount (46d)}                   & \textbf{0.9175} &	  8,613       \\ \hline
\textbf{pehash (40d)}                       & 0.8867 &	  11,835     \\ \hline
%\textbf{ssdeep (40d)}                       & 0.8798 &	  12549       \\ \hline
% \textbf{MFCC (20d)}                         & 0.898094599770203 &	  10643       \\ \hline
% \textbf{Melspectrogram (128d)}              & 0.906970509383378 &     9716	        \\ \hline
% \textbf{Chroma-STFT (12d)}                  & 0.8743010340865569&     13128	       \\ \hline
\textbf{MFCCe (320d)}                        & 0.9032 &     10,107	        \\ \hline
%\textbf{Melspectrogram-expanded (384d)}     & 0.9116717732669475&     9225	        \\ \hline
% \textbf{Chroma-STFT-expanded (312d)}        & 0.8851972424358483&     11990	       \\ \hline
\end{tabular}
\label{tbl:error_numbers-MaleX-full}
\vspace{-20pt}
\end{table}

\begin{table}[b]%[!htbp]
\centering
\caption{Error analysis for feature sets on MaleX.}
\vspace{-5pt}
\footnotesize
\begin{tabular}{|c|c|c|c|c|}
\hline
\textbf{Error}  & \textbf{gist} & \textbf{instrcount} & \textbf{pehash} & \textbf{MFCCe} \\ 
\hline
\textbf{gist}     		        & -       & 7,087   & 5,601   & 5,335  \\ \hline
\textbf{instrcount}     	    & 4,089    & -      & 3,804   & 4,227   \\ \hline
\textbf{pehash}     		    & 5,825    & 7,026    & -       & 6,234   \\ \hline
\textbf{MFCCe}     	            & 3,831    & 5,721   & 4,506   & -    \\ \hline
\end{tabular}
\label{tbl:ea_mat-MaleX-full}
% \vspace{-5pt}
\end{table}

% \begin{table}[b]%[!htbp]
% \centering
% % \tiny
% % \setlength\tabcolsep{.5pt} % default value: 6pt
% \begin{tabular}{|c|c|c|c|c|}
% \hline
% \textbf{Error}  & \textbf{gist} & \textbf{instrcount} & \textbf{pehash} & \textbf{MFCCe} \\ 
% \hline
% \textbf{gist}     		        & 0	  & 0.61   & 0.48	& 0.46   \\ \hline % div by 11611
% \textbf{instrcount}     	        & 0.47  & 0    & 0.44	 & 49    \\ \hline % div by 8613
% \textbf{pehash}     		        & 0.51 & 0.62 & 0 & 0.55  \\ \hline % div by 11385
% \textbf{MFCCe}     	            & 0.38   & 0.56  & 0.44	 & 0	   \\ \hline % div by 10107
% \end{tabular}
% \caption{Normalized Error analysis for features on the full Malex dataset.}
% \label{tbl:en_mat_full}
% \vspace{-5pt}
% \end{table}

% \begin{table}[b]%[!htbp]
% \centering
% % \tiny
% % \setlength\tabcolsep{.5pt} % default value: 6pt
% \begin{tabular}{|c|c|c|c|c|}
% \hline
% \textbf{Error}  & \textbf{gist} & \textbf{instrcount} & \textbf{pehash} & \textbf{MFCCe} \\ 
% \hline
% \textbf{gist}     		        & 0	    & 0.67   & 0.64	 & 0.59   \\ \hline % div by 11611
% \textbf{instrcount}     	        & 0.67  & 0      & 0.66	 & 0.66    \\ \hline % div by 8613
% \textbf{pehash}     		        & 0.64  & 0.66   & 0     & 0.64  \\ \hline % div by 11385
% \textbf{MFCCe}     	            & 0.59  & 0.66   & 0.64	 & 0	   \\ \hline % div by 10107
% \end{tabular}
% \caption{Error analysis for features on the full Malex dataset.}
% \label{tbl:en_mat_full}
% \vspace{-5pt}
% \end{table}

\begin{table}[b]%[!htbp]
\centering
\caption{$JFS$ score matrix comparing orthogonality of different feature sets on MaleX -- normalization factors are derived from the last column of Table~\ref{tbl:error_numbers-MaleX-full}.}
\vspace{-5pt}
\footnotesize
\begin{tabular}{|c|c|c|c|c|}
\hline
\textbf{$JFS$ Score}    & \textbf{gist} & \textbf{instrcount} & \textbf{pehash} &  \textbf{MFCCe} \\ 
\hline
\textbf{gist}     		     & -     & 0.67   & 0.64      & 0.59 \\ \hline  
\textbf{instrcount}     	 & 0.67    & -    & 0.65      & 0.67 \\ \hline   
\textbf{pehash}     		 & 0.64    & 0.65   & -       & 0.64 \\ \hline   
%\textbf{ssdeep}     		 & 0.72    & 0.72   & 0.72   & 0    & 0.69 \\ \hline   
\textbf{MFCCe}               & 0.59    & 0.67   & 0.64    & - \\ \hline
% \hline
% \textbf{$JFS$ Score}    & \textbf{gist} & \textbf{instrcount} & \textbf{pehash} &  \textbf{mfcce} \\ 
% \hline
% \textbf{gist}     		     & 0     & 0.74   & 0.74      & 0.69 \\ \hline  
% \textbf{instrcount}     	 & 0.74    & 0    & 0.72      & 0.75 \\ \hline   
% \textbf{pehash}     		 & 0.74    & 0.72   & 0       & 0.72 \\ \hline   
% %\textbf{ssdeep}     		 & 0.72    & 0.72   & 0.72   & 0    & 0.69 \\ \hline   
% \textbf{MFCCe}               & 0.69    & 0.75   & 0.72    & 0 \\ \hline
\end{tabular}
\label{tbl:JFS_mat-MaleX-full}
\vspace{-5pt}
\end{table}

\subsection{Large scale analysis on MaleX dataset}

%During the current period we performed  all our experiments on the bigger MaleX dataset, which include
We validate our approach on the bigger MaleX~\cite{mohammed2021malware} dataset, which contains 179,725 benign samples and 864,669 malicious samples (both packed and unpacked) pertaining to Windows OS. 
%Here, we exploit local feature variations and average the features for uniformly divided parts of the binary files, and then compare the orthogonality scores.  
%\subsection*{Results accounting for local feature variations}
%We used the entire MaleX dataset for this experiment and the results are shown in Table~\ref{table:exp-dim}.
We follow the same procedure as the previous experiment and and the results are shown in Table~\ref{table:exp-dim}.
% The errors corresponding to different features we have used for malware classification in the balanced MaleX dataset (179725 benign and malicious samples) when using random-forest classifier is listed in Table~\ref{tbl:error_numbers-MaleX-bal}.
% %\pagebreak
% The error analysis matrix ($E$) for these features is shown in Table~\ref{tbl:ea_mat-MaleX-bal}.
% The $JFS$ score for each pair of features is shown in Table~\ref{tbl:JFS_mat-MaleX-bal}.
The errors corresponding to different feature sets of the MaleX dataset (179,725 benign and 864,669 malicious samples) when using random-forest classifier is listed in Table~\ref{tbl:error_numbers-MaleX-full} (results were similar for other classifiers too).
The error analysis matrix ($E$) for these feature sets is shown in Table~\ref{tbl:ea_mat-MaleX-full}.
The $JFS$ score for each pair of feature sets is shown in Table~\ref{tbl:JFS_mat-MaleX-full}.
As we can see from these results, our approach is scalable even on a dataset that is 66 times larger (1 million samples). 

% \begin{itemize}
%     \item Varying bytes per audio sample, wherein we created audio signal samples using more than 2 or 4 bytes per sample (instead of 1). This turned out to be not helpful. 
%     \item Exploiting local feature variations, we averaged features for uniformly divided parts of the binary files. This improved classification accuracy for many features. 
%     \item Orthogonality scores comparing the new audio features with previously evaluated features.
% \end{itemize}

% \subsection{Results with different bytes per sample}
% Results for 1, 2, 4 byte(s) per sample are provided in Tables \ref{table:8bps}, \ref{table:16bps} and \ref{table:32bps}, respectively.
% We used the full MaleX dataset for these experiments, which contain 179725 benign samples and 864669 malicious samples. Similar to smalldata changing the bytes per sample did not help.
% \vspace{-5pt}
\section{Discussion}

\noindent \textbf{Observation 1 -- Orthogonality test helps in better fusion of feature sets:} The orthogonality test checks if new feature sets help in identifying malware samples that are not already identified by existing features; thus an analyst can decide a go/no-go on a new feature set.
Now, the question could be asked if we could combine the predictions of the feature sets that passed the test and build another meta-classifier. 
We performed this experiment for the feature sets in Table~\ref{tab:acc-comp} with the Nearest Neighbor classifier and obtained a fusion accuracy of 99.98\% (0.3\% improvement over the feature with highest accuracy and 5\% improvement over the feature with lowest accuracy).
This shows that there is scope for improvement after passing the orthogonality test and it could be left to the analyst to decide if the analyst would like to keep the individual feature predictions or the fused prediction result.  

\noindent \textbf{Observation 2 -- The choice of machine learning (ML) classifiers has a minimal impact on performance:} From our experimental results both on the small dataset and the MaleX dataset, we observe that the choice of our ML classifier has negligible impact on the accuracy of the features.
Hence, this choice can be left to an analyst depending on the use case. 
For example, in certain defense/military applications, use of Nearest Neighbor classifier is preferable since the prediction can easily be mapped to the nearest neighbor and the distance to it, thus providing some explainability and confidence. 

\noindent \textbf{Observation 3 -- OMD framework is generalizable to other features:} We primarily focused on feature sets that are computed fast (milliseconds to seconds) such as static and signal processing (audio and image) features. We picked these features as a proof of concept for our overall framework of error analysis and orthogonality test. 
This does not stop an analyst from adding complex feature sets based on dynamic analysis, which could take longer time to compute (minutes to hours), but could easily be added to our framework, once the predictions from these features have been completed.

\noindent \textbf{Observation 4 -- $JFS$ is sensitive to class imbalance:}
The $JFS$ metric used to determine orthogonality between a pair of feature sets shows better resilience when the class distribution is balanced.
%is not fully resilient in class imbalanced settings. 
For example, Table~\ref{tbl:JFS_mat-MaleX-subb} shows the results for a balanced subset of MaleX dataset that contains 179,725 benign and 179,725 randomly sampled malware samples from the full Malex dataset.
These results better support the orthogonality claims made for various pairs of the feature sets
% \hl{seem trustworthy} 
similar to that of the somewhat-balanced small dataset (Table~\ref{tbl:JFS_mat}). 
However, these results are slightly different when compared to that of the entire MaleX dataset (Table~\ref{tbl:JFS_mat-MaleX-full}) which has more malware samples. 
This suggests that while using the $JFS$ metric for orthogonality measure, it is worthwhile to consider a more balanced data setup which can be obtained using any random sampling strategy that maintains the data diversity.
% \hl{Optional TODO:  we may want to point to data diversity if space permits} 

\vspace{-5pt}
\begin{table}[ht]%[!htbp]
\centering
\caption{$JFS$ score matrix comparing orthogonality of different feature sets on a \textit{balanced} subset of MaleX dataset.}
\vspace{-5pt}
\footnotesize
\begin{tabular}{|c|c|c|c|c|}
\hline
\textbf{$JFS$ Score}    & \textbf{gist} & \textbf{instrcount} & \textbf{pehash} &  \textbf{MFCCe} \\ 
\hline
\textbf{gist}     		     & -     & 0.75   & 0.76      & 0.66 \\ \hline  
\textbf{instrcount}     	 & 0.75    & -    & 0.78      & 0.75 \\ \hline   
\textbf{pehash}     		 & 0.76    & 0.78   & -       & 0.75 \\ \hline   
%\textbf{ssdeep}     		 & 0.72    & 0.72   & 0.72   & 0    & 0.69 \\ \hline   
\textbf{MFCCe}               & 0.66    & 0.75   & 0.75    & - \\ \hline
% \hline
% \textbf{$JFS$ Score}    & \textbf{gist} & \textbf{instrcount} & \textbf{pehash} &  \textbf{mfcce} \\ 
% \hline
% \textbf{gist}     		     & 0     & 0.74   & 0.74      & 0.69 \\ \hline  
% \textbf{instrcount}     	 & 0.74    & 0    & 0.72      & 0.75 \\ \hline   
% \textbf{pehash}     		 & 0.74    & 0.72   & 0       & 0.72 \\ \hline   
% %\textbf{ssdeep}     		 & 0.72    & 0.72   & 0.72   & 0    & 0.69 \\ \hline   
% \textbf{MFCCe}               & 0.69    & 0.75   & 0.72    & 0 \\ \hline
\end{tabular}
\label{tbl:JFS_mat-MaleX-subb}
\vspace{-10pt}
\end{table}
% \section{Todos}
% TODO- add figure from report for pipeline: Satish--title looks good on further thinking... we are checking orthogonality of all various features, not just audio. 

% TODO: should we discuss adversarial ML?

% TODO: KNN VS NN VS RF

% %TODO- add fusion study

% TODO- add that we can add other features to the audio treatment (i,e, other features can be represented in the spectra)

% TODO-Performance discussion 
% --(a) timing info (Rough idea ..ms/sec/min/hours)
% --(b) lossiness? 

% TODO-impact of Packing

% todo: Add observation1, 2, 3 as in the image paper

%%%%%%%%%%%%%%%%%%%%%%%%%%%%%%%%%%%%%%%%%%%%%%%%%%%%%%%%%%%%%%%%%%%%%%
% \vspace{-6pt}
\section{Conclusion and Future Work}
%\vspace{-5pt}

% \bsm{looks like much work is still needed to fill this in. Who is in charge of this paper?}

In this paper, we proposed a novel framework for orthogonal malware detection (OMD) using features based on audio, image, and static analysis.
We showed that audio descriptors are effective in detecting and classifying malware, and the predictions made on the audio descriptors are orthogonal to the predictions made on other feature sets.
We evaluated our approach using an orthogonality metric which quantifies how orthogonal a new feature set is with respect to other feature sets. 
Thus our overall framework and approach can be easily extended and adapted to allow new features and detection methods to be added.
In future, we will develop a malware informatics engine which integrates signal/image based malware techniques with standard security methods and provide insights both from a signal/image and security point of view.

% \vspace{-5pt}

%%%%%%%%%%%%%%%%%%%%%%%%%%%%%%%%%%%%%%%%%%%%%%%%%%%%%%%%%%%%%%%%%%%%%%%%

%\section*{References}

% Please number citations consecutively within brackets \cite{b1}. The 
% sentence punctuation follows the bracket \cite{b2}. Refer simply to the reference 
% number, as in \cite{b3}---do not use ``Ref. \cite{b3}'' or ``reference \cite{b3}'' except at 
% the beginning of a sentence: ``Reference \cite{b3} was the first $\ldots$''

% Number footnotes separately in superscripts. Place the actual footnote at 
% the bottom of the column in which it was cited. Do not put footnotes in the 
% abstract or reference list. Use letters for table footnotes.

% Unless there are six authors or more give all authors' names; do not use 
% ``et al.''. Papers that have not been published, even if they have been 
% submitted for publication, should be cited as ``unpublished'' \cite{b4}. Papers 
% that have been accepted for publication should be cited as ``in press'' \cite{b5}. 
% Capitalize only the first word in a paper title, except for proper nouns and 
% element symbols.

% For papers published in translation journals, please give the English 
% citation first, followed by the original foreign-language citation \cite{b6}.

\bibliographystyle{IEEEtran}
% \bibliography{bib/malware,bib/mal,bib/mc,bib/specific,bib/mchicss}

\begin{thebibliography}{10}
\providecommand{\url}[1]{#1}
\csname url@samestyle\endcsname
\providecommand{\newblock}{\relax}
\providecommand{\bibinfo}[2]{#2}
\providecommand{\BIBentrySTDinterwordspacing}{\spaceskip=0pt\relax}
\providecommand{\BIBentryALTinterwordstretchfactor}{4}
\providecommand{\BIBentryALTinterwordspacing}{\spaceskip=\fontdimen2\font plus
\BIBentryALTinterwordstretchfactor\fontdimen3\font minus
  \fontdimen4\font\relax}
\providecommand{\BIBforeignlanguage}[2]{{%
\expandafter\ifx\csname l@#1\endcsname\relax
\typeout{** WARNING: IEEEtran.bst: No hyphenation pattern has been}%
\typeout{** loaded for the language `#1'. Using the pattern for}%
\typeout{** the default language instead.}%
\else
\language=\csname l@#1\endcsname
\fi
#2}}
\providecommand{\BIBdecl}{\relax}
\BIBdecl

\bibitem{nataraj2011malware}
L.~Nataraj, S.~Karthikeyan, G.~Jacob, and B.~Manjunath, ``Malware images:
  visualization and automatic classification,'' in \emph{Proc. International
  symposium on visualization for cyber security}.\hskip 1em plus 0.5em minus
  0.4em\relax ACM, 2011, p.~4.

\bibitem{nataraj2016spam}
L.~Nataraj and B.~Manjunath, ``Spam: Signal processing to analyze malware
  [applications corner],'' \emph{IEEE Signal Processing Magazine}, vol.~33,
  no.~2, pp. 105--117, 2016.

\bibitem{vinayakumar2019robust}
R.~Vinayakumar, M.~Alazab, K.~Soman, P.~Poornachandran, and S.~Venkatraman,
  ``Robust intelligent malware detection using deep learning,'' \emph{IEEE
  Access}, vol.~7, pp. 46\,717--46\,738, 2019.

\bibitem{gibert2020rise}
D.~Gibert, C.~Mateu, and J.~Planes, ``The rise of machine learning for
  detection and classification of malware: Research developments, trends and
  challenges,'' \emph{Journal of Network and Computer Applications}, 2020.

\bibitem{mohammed2021malware}
T.~M. Mohammed, L.~Nataraj, S.~Chikkagoudar, S.~Chandrasekaran, and
  B.~Manjunath, ``Malware detection using frequency domain-based image
  visualization and deep learning,'' in \emph{Proceedings of the 54th Hawaii
  International Conference on System Sciences}, 2021, p. 7132.

\bibitem{nguyen2018auto}
M.~H. Nguyen, D.~Le~Nguyen, X.~M. Nguyen, and T.~T. Quan, ``Auto-detection of
  sophisticated malware using lazy-binding control flow graph and deep
  learning,'' \emph{Computers \& Security}, vol.~76, pp. 128--155, 2018.

\bibitem{ma2019combination}
Z.~Ma, H.~Ge, Y.~Liu, M.~Zhao, and J.~Ma, ``A combination method for android
  malware detection based on control flow graphs and machine learning
  algorithms,'' \emph{IEEE access}, vol.~7, pp. 21\,235--21\,245, 2019.

\bibitem{zhang2019classification}
H.~Zhang, X.~Xiao, F.~Mercaldo, S.~Ni, F.~Martinelli, and A.~K. Sangaiah,
  ``Classification of ransomware families with machine learning based on n-gram
  of opcodes,'' \emph{Future Generation Computer Systems}, 2019.

\bibitem{namanya2020similarity}
A.~P. Namanya, I.~U. Awan, J.~P. Disso, and M.~Younas, ``Similarity hash based
  scoring of portable executable files for efficient malware detection in
  iot,'' \emph{Future Generation Computer Systems}, 2020.

\bibitem{rezaei2020efficient}
T.~Rezaei and A.~Hamze, ``An efficient approach for malware detection using pe
  header specifications,'' in \emph{2020 6th International Conference on Web
  Research (ICWR)}.\hskip 1em plus 0.5em minus 0.4em\relax IEEE, 2020, pp.
  234--239.

\bibitem{zhang2016malware}
J.~Zhang, Z.~Qin, H.~Yin, L.~Ou, S.~Xiao, and Y.~Hu, ``Malware variant
  detection using opcode image recognition with small training sets,'' in
  \emph{2016 25th International Conference on Computer Communication and
  Networks (ICCCN)}.\hskip 1em plus 0.5em minus 0.4em\relax IEEE, 2016, pp.
  1--9.

\bibitem{farrokhmanesh2016novel}
M.~Farrokhmanesh and A.~Hamzeh, ``A novel method for malware detection using
  audio signal processing techniques,'' in \emph{2016 Artificial Intelligence
  and Robotics (IRANOPEN)}.\hskip 1em plus 0.5em minus 0.4em\relax IEEE, 2016,
  pp. 85--91.

\bibitem{sharma2018image}
P.~Sharma and A.~Raglin, ``Image-audio encoding for information camouflage and
  improving malware pattern analysis,'' in \emph{2018 17th IEEE International
  Conference on Machine Learning and Applications (ICMLA)}.\hskip 1em plus
  0.5em minus 0.4em\relax IEEE, 2018, pp. 1059--1064.

\bibitem{azab2020msic}
A.~Azab and M.~Khasawneh, ``Msic: malware spectrogram image classification,''
  \emph{IEEE Access}, vol.~8, pp. 102\,007--102\,021, 2020.

\bibitem{mercaldo2021audio}
F.~Mercaldo and A.~Santone, ``Audio signal processing for android malware
  detection and family identification,'' \emph{Journal of Computer Virology and
  Hacking Techniques}, vol.~17, no.~2, pp. 139--152, 2021.

\bibitem{kolter2006learning}
J.~Z. Kolter and M.~A. Maloof, ``Learning to detect and classify malicious
  executables in the wild,'' \emph{JLMR}, 2006.

\bibitem{lakhotia2013vilo}
A.~Lakhotia, A.~Walenstein, C.~Miles, and A.~Singh, ``Vilo: a rapid learning
  nearest-neighbor classifier for malware triage,'' \emph{Journal of Computer
  Virology and Hacking Techniques}, vol.~9, no.~3, 2013.

\bibitem{wicherski2009pehash}
G.~Wicherski, ``pehash: A novel approach to fast malware clustering,'' in
  \emph{2nd USENIX Workshop on LEET}, 2009.

\bibitem{nataraj11comparative}
L.~Nataraj, V.~Yegneswaran, P.~Porras, and J.~Zhang, ``A comparative assessment
  of malware classification using binary texture analysis and dynamic
  analysis,'' in \emph{AISec}, Oct 2011.

\bibitem{mcfee2015librosa}
B.~McFee, C.~Raffel, D.~Liang, D.~P. Ellis, M.~McVicar, E.~Battenberg, and
  O.~Nieto, ``librosa: Audio and music signal analysis in python,'' in
  \emph{Proceedings of the 14th Python in Science Conference}, 2015.

\bibitem{schorkhuber2010constant}
C.~Sch{\"o}rkhuber and A.~Klapuri, ``Constant-q transform toolbox for music
  processing,'' in \emph{Proceedings of the International Conference on Music
  Information Retrieval (ISMIR)}, 2010.

\bibitem{muller2005audio}
M.~M{\"u}ller, F.~Kurth, and M.~Clausen, ``Audio matching via chroma-based
  statistical features.'' in \emph{Proceedings of the International Conference
  on Music Information Retrieval (ISMIR)}, 2005, pp. 288--295.

\bibitem{rabiner1993fun}
L.~R. Rabiner and B.~H. Juang, \emph{Fundamentals of Speech Recognition}.\hskip
  1em plus 0.5em minus 0.4em\relax Prentice Hall, 1993.

\bibitem{branco2012scientific}
R.~R. Branco, G.~N. Barbosa, and P.~D. Neto, ``Scientific but not academical
  overview of malware anti-debugging, anti-disassembly and anti-vm
  technologies,'' \emph{Blackhat USA}, vol.~12, 2012.

\end{thebibliography}

% \bibitem{b1} G. Eason, B. Noble, and I. N. Sneddon, ``On certain integrals of Lipschitz-Hankel type involving products of Bessel functions,'' Phil. Trans. Roy. Soc. London, vol. A247, pp. 529--551, April 1955.
% \bibitem{b2} J. Clerk Maxwell, A Treatise on Electricity and Magnetism, 3rd ed., vol. 2. Oxford: Clarendon, 1892, pp.68--73.
% \bibitem{b3} I. S. Jacobs and C. P. Bean, ``Fine particles, thin films and exchange anisotropy,'' in Magnetism, vol. III, G. T. Rado and H. Suhl, Eds. New York: Academic, 1963, pp. 271--350.
% \bibitem{b4} K. Elissa, ``Title of paper if known,'' unpublished.
% \bibitem{b5} R. Nicole, ``Title of paper with only first word capitalized,'' J. Name Stand. Abbrev., in press.
% \bibitem{b6} Y. Yorozu, M. Hirano, K. Oka, and Y. Tagawa, ``Electron spectroscopy studies on magneto-optical media and plastic substrate interface,'' IEEE Transl. J. Magn. Japan, vol. 2, pp. 740--741, August 1987 [Digests 9th Annual Conf. Magnetics Japan, p. 301, 1982].
% \bibitem{b7} M. Young, The Technical Writer's Handbook. Mill Valley, CA: University Science, 1989.
% \end{thebibliography}
% \vspace{12pt}
% \color{red}
% IEEE conference templates contain guidance text for composing and formatting conference papers. Please ensure that all template text is removed from your conference paper prior to submission to the conference. Failure to remove the template text from your paper may result in your paper not being published.

\end{document}